# Creating and Operating Areas With Reduced Electromagnetic Field Exposure Thanks to Reconfigurable Intelligent Surfaces

Dinh-Thuy Phan-Huy, Yohann Bénédic*, Sebastien Herraiz Gonzalez*, Philippe Ratajczak**
Orange Innovation
Châtillon, *Belfort, **Sophia-Antipolis, France,
{dinhthuy.phanhuy, yohann.benedic, sebastien.herraiz, philippe.ratajczak}@orange.com

*Abstract*— **Mobile network operators must provide a target quality of service (QoS) within a target coverage area. Each generation of networks from the 2nd to the 5th has reached higher QoS targets and coverage area sizes. However, the deployment of new networks is sometimes challenged by Electromagnetic Field (EMF) exposure constraints. In this paper, to take into account these constraints, we assess the novel and recent concept of "Reduced EMF Exposure Area". Such an area would be created and operated by a mobile network operator upon the request of its customers. In such an area, customers keep enjoying high data rate internet access while getting a reduced EMF exposure. To ensure this EMF exposure reduction, we propose to deploy Reconfigurable Intelligent Surfaces (RIS) connected to the mobile network and exploit a joint RIS-M-MIMO uplink beamforming scheme. We use our ray-tracing-based simulation tool to visualize and characterize the Reduced EMF Exposure Area in a challenging environment in terms of propagation. Our simulations show that an operator can create and operate such an area under the condition that it carefully places the RIS in the environment.**

I. INTRODUCTION

The deployment of the 5th generation (5G) of networks is already challenged by Electromagnetic field (EMF) exposure constraints [1]. One of the means to reduce EMF exposure is the use of reconfigurable intelligent surfaces (RIS) [2-4], which are arrays of resonators that re-radiate impinging waves, in a controlled manner (with a chosen phase-shift and/or amplitude). Indeed, in [5], techniques have been proposed to control the EMF exposure around and close to a massive multiple input multiple output (MMIMO) [6] base station (BS) implementing adaptive beamforming, based on RIS. In parallel, [7][8] optimum joint RIS and BF schemes have been derived to reduce the self-EMF exposure caused by the mobile device on its user, when it is transmitting data in the uplink (UL). Recently, [9-11] has proposed the new concept of performance "reduced EMF exposure area, where the network operator offers as a service to its customer the possibility to get reduced EMF exposure for the same target Quality of service (QoS), by shaping the propagation with a RIS. Finally, in [12], the first deployment and planning methods of RISs have been designed, using ray-tracing based simulations to take into account realistic models of the propagation channels between the BS, the user equipment (UE), and the RIS.

In this paper, we propose to assess the validity of this recent and new concept of "reduced EMF exposure Area". In such an area, for a given target service and a given target data rate, the self-exposure to EMF is reduced by the network operator, thanks to the utilization of a RIS. We propose a visualization of such an area in a challenging environment regarding propagation and self-exposure. To that aim, we evaluate numerically the reduction in self-EMF exposure that can be obtained from the introduction of a "giant" square RIS of 100 by 100 unit cells, with continuous phase shifting capabilities, as the 14 by 14 prototype successfully used for experiments in [13][14]. Such "giant" RIS enables us to get a flavor of the upper bound performance. As in [12], we will use a three-dimensional (3D) ray-tracing simulation tool (the Orange internal STARLIGHT® tool) to capture the impact of the propagation faithfully.

The paper is organized as follows: Section II presents our system model. Section III presents our deployment scenario and simulation results and Section IV concludes this paper.

II. SYSTEM MODEL

We consider the UL communication between one UE equipped with one transmit antenna and a M-MIMO BS with $N$ receive antennas. We consider one RIS with $K$ unit cells with continuous phase-shifting capabilities [13][14].

We assume that the RIS has sounding capabilities and that each unit cell is capable of estimating the channel between the UE and itself. More precisely, we assume that every radio frame is split into several time slots and that the UE sends UL sounding pilots (i.e. pilots intended to the RIS); during the first time slot, and then sends UL data along with demodulation pilots (i.e.

pilots intended to the BS) in the next time slots of the same radio frame. During the first time slot of the radio frame, the RIS estimates the channel based on the UL sounding pilots and computes its reflection beamforming weights. The beamforming weights are computed following the method in [5]. During the next time slots of the same radio frame, the BS estimates the UL channel based on the demodulation pilots and equalizes the received UL data using maximum ratio combining (MRC).

We also assume that the UE is not moving or moving slowly and that the channel is coherent for at least one radio frame. Such an assumption is a reasonable assumption for an indoor deployment propagation scenario for instance.

We denote by $\mathbf{h} \in \mathbb{C}^{N\times 1}$ the propagation channel vector between the UE and the BS, in absence of the RIS. We denote by $\mathbf{w} \in \mathbb{C}^{K\times 1}$ the propagation channel vector between the UE and the RIS. We denote by $\mathbf{q} \in \mathbb{C}^{K\times N}$ the propagation channel vector between the RIS and the BS. We denote by $\mathbf{b} \in \mathbb{C}^{K\times K}$ the diagonal matrix of the reflection beamforming weights of the RIS. We denote by $\mathbf{g} \in \mathbb{C}^{N\times 1}$ the propagation channel between the UE and the BS, in presence of the RIS. With these notations and assumptions, $\mathbf{b}$ is given by [5]:

$$\mathbf{b}(k,k) = \frac{\mathbf{w}^h(k,k)}{|\mathbf{w}^h(k,k)|}, k \in [\![1;K]\!] \quad (1)$$

where $(.)^h$ is the Hermitian operation and $|\mathbf{w}^h(k,k)|$ is the module of $\mathbf{w}^h(k,k)$. As a consequence, $\mathbf{g}$ is given by:

$$\mathbf{g} = \mathbf{wbq} + \mathbf{h}, \text{ in the presence of RIS, and} \quad (2)$$

$$\mathbf{g} = \mathbf{h}, \text{ in the absence of RIS.}$$

We denote by $N_o$, $W$ and $N_f$, the noise power spectral density, the bandwidth and the BS receiver noise figure, respectively. We denote by $P^t$ the UE transmit power. With these notations, the signal to noise ratio (SNR) at the output of the MRC receiver is given by:

$$SNR = \mathbf{g}^h\mathbf{g}\frac{P^t}{N_fN_oW}, \quad (3)$$

where $\mathbf{g}^h\mathbf{g} \in \mathbb{R}^+$ is the UL equivalent channel after MRC equalization.

We assume that an UL closed loop power control mechanism is implemented to ensure that the UE transmit power $P^t$ (and thus the user self-exposure) is equal to the value just necessary for the given service. We assess the $P^t$ for a target service with a target rate $R$. The target spectral efficiency $SE$ is given by $R/W$, and the corresponding target SNR $SNR_{target}$ satisfies the Shannon capacity formula: $SE = log2(1 + SNR_{target})$. We denote by $P^{target}$ the UE transmit power that would be required at the UE side, to meet the target SNR:

$$P^{target} = \frac{SNR_{target}N_fN_oW}{\mathbf{g}^h\mathbf{g}} \quad (4)$$

The UE actual transmit power $P^t$ is bounded by the maximum and minimum values (due to the UE power amplifier limitations), $P^{min}$ and $P^{max}$, respectively. Therefore, if $P^{target} > P^{max}$, then, the UE is considered out of coverage for this specific service and served in the uplink direction otherwise. Finally, if $P^{target} < P^{min}$, then the UE transmit at $P^t = P^{min}$.

The larger is $P^t$ and the larger is the self-exposure of the user. We will therefore assess $P^t$ to evaluate the impact of the introduction of the RIS on the EMF self-exposure of the user.

III. DEPLOYMENT SCENARIO AND SIMULATION RESULTS

This current section presents the studied deployment scenario and simulation results. More precisely, in sub-section A, we describe the deployment and propagation scenario and our simulation methodology. Then, in Sub-Section B, we provide a quantitative evaluation of the improvement of the equivalent channel gain (after MRC) brought by the RIS. Finally, in Sub-Section C, we visualize the reduced self-exposure areas concept.

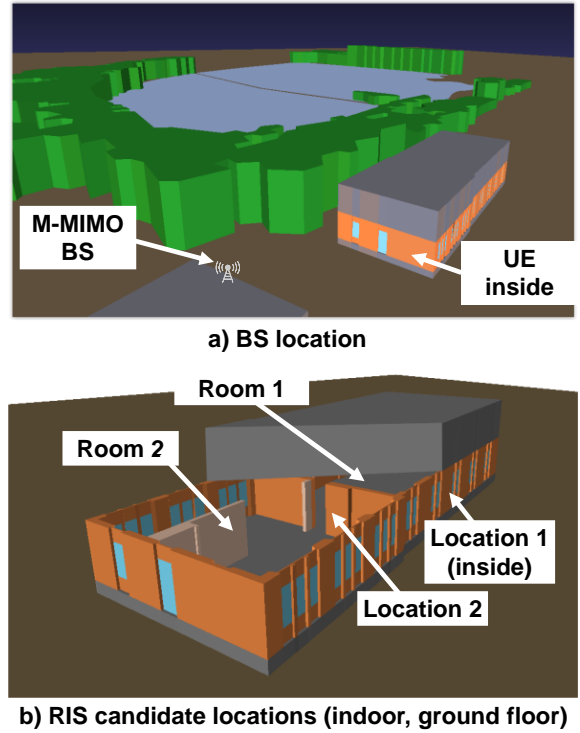

a) BS location

b) RIS candidate locations (indoor, ground floor)

Fig. 1. BS and RIS locations

*A. Deployment scenario and simulation methodology*

We study a typically challenging scenario in terms of self-EMF exposure. We consider a simple indoor-to-outdoor deployment scenario, where a UE is located on the ground floor inside a building, with two rooms, and served in the UL by an MMIMO macro BS deployed outside, on the roof of another building (as illustrated by Fig. 1-a)) facing the first building. Such a scenario is a typically highly challenging scenario in terms of propagation and self-EMF exposure. Indeed, due to the strong path loss, when the UE is in the farthest room from the BS, it may transmit at a very high power to reach the target SNR, or be even out of coverage.

The carrier frequency $f$ is set to 3.7 GHz, which is one of the carrier frequencies with which MMIMO 5G BSs are currently deployed. Two rooms are considered: room 1 is the farthest from the BS and room 2 is the closest one. The BS is equipped with $N = 32$ antennas (4 columns of 8 dual-polarized antenna elements spaced by half a wavelength in the vertical domain, and 0.8 wavelength in the horizontal domain.

One RIS only is deployed, however, as in [8] several candidate locations are studied: location 1 inside room 1 (with a good propagation from itself to the BS) and location 2 inside room 2 (with a very poor propagation from itself to the BS) are considered (as illustrated by Fig. 1-b)). The RIS is "giant" a square array of 100 by 100 elements, i. e. $k = 10,000$ unit cells, spaced by 20 cm (i.e. approximatively a quarter of a wavelength) as in the RIS prototype of [13][14]. Such "giant" RIS enables obtaining a flavor of the upper bound of the performance of such a system. Indeed, smaller RIS(s) in size are expected to exhibit poorer performance. For each location, the RIS is standing up upon the ground (with its center at 1-meter height above the floor) and against a wall.

TABLE I. PARAMETERS VALUES

| Parameter | Value | Unit |
|---|---|---|
| $N_0$ | -174 [15] | dBm/Hz |
| $N_f$ | 5 [15] | dB |
| $f$ | 3.7 | GHz |
| $\lambda$ | ~20 | cm |
| $K$ | 10,000 | Unit cells |
| $N$ | 32 | Antenna elements |
| $P^{max}$ | 23 [16] | dBm |
| $P^{min}$ | 0 | dBm |
| Target service | | |
| $R$ | 30 | kbps |
| $W$ | 30 | kHz |
| $rate$ | 30 | Kbps |
| SE | 1 | Bits/s/Hz (QPSK) |
| $SNR_{target}$ | 0 | dB |

Figure 2-a) and 2-b) illustrate the rays between the center of the RIS and the 32 elements of the BS, for location 1 and location 2 of the RIS, respectively. One can observe that for both RIS locations, rays enter the building through windows facing the BS. More rays hit location 2, thanks to reflections by the walls of room 2. Location 1 suffers from a stronger keyhole effect and benefits from nearly no reflections inside room 1. Location 2 of the RIS is therefore expected to have a significantly better propagation to the BS than the RIS1, for two reasons: 1) it is closer to the BS and shaded by fewer walls and 2) it benefits from more reflections on walls inside the room 2.

To assess the self-exposure reduction brought by the introduction of a RIS, we propose to compute numerically, maps of the UE transmit power for a given target service. As the chosen propagation scenario is highly challenging, we propose to assess a low data rate of 30 kbps, which is a typical rate for voice service (the detailed parameters for this service are listed in Table I). More precisely, for each location of the UE in the building ground floor, at 1 meter height above the floor (which is the typical height for a UE on a table, a desk or in the hand of a user, even when it is used for voice communication for instance with a headset), we generate the propagation channels **h, q** and **w** using Orange internal 3D ray-tracing tool STARLIGHT® and using antenna diagrams of the same unit cells as in the prototypes of [13][14] and the antenna diagrams of the MMIMO BS antenna elements (patch antennas). Then, for each location of the UE, we compute **b** and derive the UE transmit power $P^t$ as described in Section II, with the parameter values settings provided in Table I.

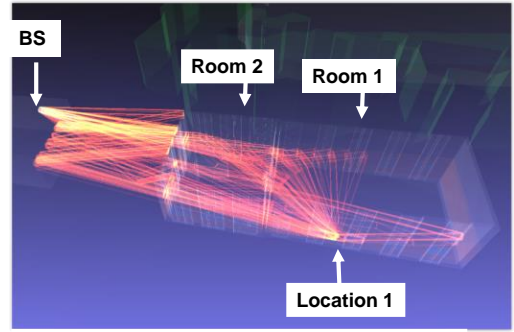

a) RIS location 1

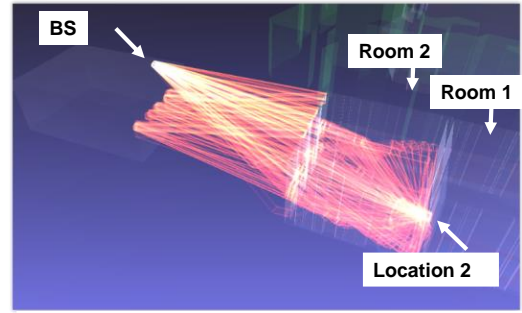

b) RIS location 2

Fig. 2. Rays between the center of the RIS and the BS 32 antenna elements for a) Location 1 and b) Location 2 of the RIS.

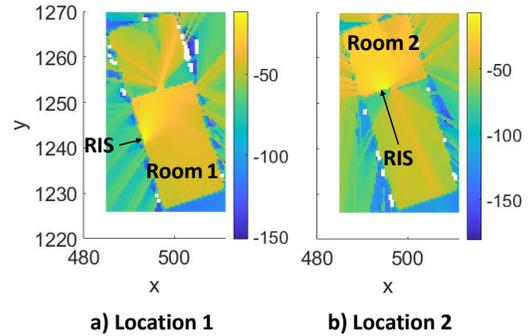

a) Location 1     b) Location 2

Fig. 3. UE to RIS channel power in dB for RIS locations 1 and 2, respectively.

In order to visualize how the RIS would behave as an "active relay" capable of transmitting, we plot in Fig. 3 a map of the UE-to-RIS channel power $\mathbf{w}^h\mathbf{w} \in \mathbb{R}^+$ for RIS location 1 and 2 respectively. If the RIS was an active relay, location 1 and location 2, would mainly cover room 1 and room 2, respectively. However, we also note that whatever its location, a RIS in one room manages to also cover (to a lower extent) the other room, in particular, thanks to reflections on neighboring walls.

## B. Equivalent channel improvement

A map of the equivalent channel gain $\mathbf{g}^h\mathbf{g} \in \mathbb{R}^+$ between the UE and the BS when RIS is absent is plotted in Fig. 4-a). One can observe a strong discrepancy between the two rooms. Room 2 is much better covered than room 1 (100 dB of difference at least). Therefore, we expect the UE transmit power (and thus the user self-exposure) to be much stronger in room 1.

A map of the improvement of $\mathbf{g}^h\mathbf{g}$ obtained thanks to the introduction of the RIS is plotted in Fig. 4-b) for location 1 and location 2 of the RIS. First of all, we can observe that only room 1 is improved. This is due to the fact that in room 2, the rays coming directly to the BS are more numerous and stronger than the rays reflected by the RIS. In general, all locations with good propagation conditions even without the RIS (i.e. all locations in room 2 and some locations in room 1), do not benefit from the introduction of the RIS, whatever the RIS position.

Surprisingly, the propagation from room 1 to the BS is better improved by the RIS when it is located in the other room (room 2) and not when it is located in the same room (room 1). Indeed, with an active relay (instead of a RIS) deploying the relay node in room 1 would be expected to be the best solution to improve room 1 coverage. However, the RIS is a passive relay, which performance strongly depends on the propagation between itself and the BS. As illustrated in Fig. 4-b), in the simulated scenario, the propagation from any location in room 2 (including location 2) is much better than any location in room 1 (including location 1), by a hundred of dBs. As the RIS is a passive relay that reflects the impinging wave, it is better to position it in a location with good propagation conditions. Also, even if the RIS is located in room 2, thanks to the reflections on other walls of room 2, it can redirect energy towards room 1, through the door between the two rooms.

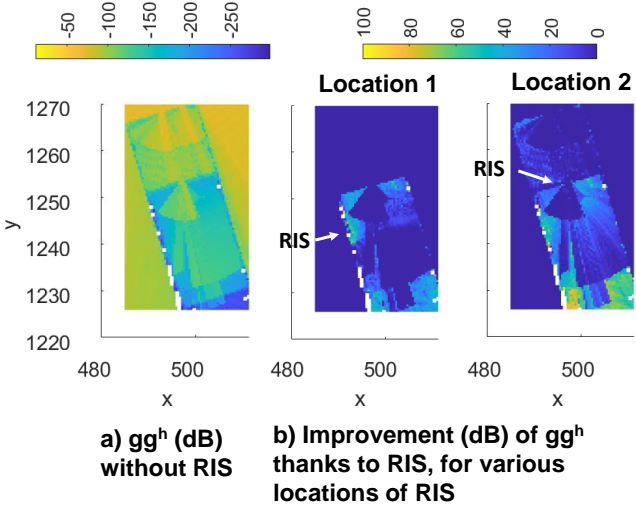

Fig. 4. UL equivalent channel gain ($\mathbf{g}^h\mathbf{g}$) in dB a) without RIS, b) with RIS at location 1 and c) RIS at location 2.

The improvement of $\mathbf{g}^h\mathbf{g}$ obtained thanks to the RIS varies a lot and can reach 100 dBs. However, this may not systematically translate into self-exposure reduction. Indeed, if, even with its maximum power and with this propagation improvement, the UE can still not reach the target SNR of the considered target service, then, the UE may remain out of coverage. The next section, will allow to visualize which area is improved, taking into account the target service and out-of-coverage locations.

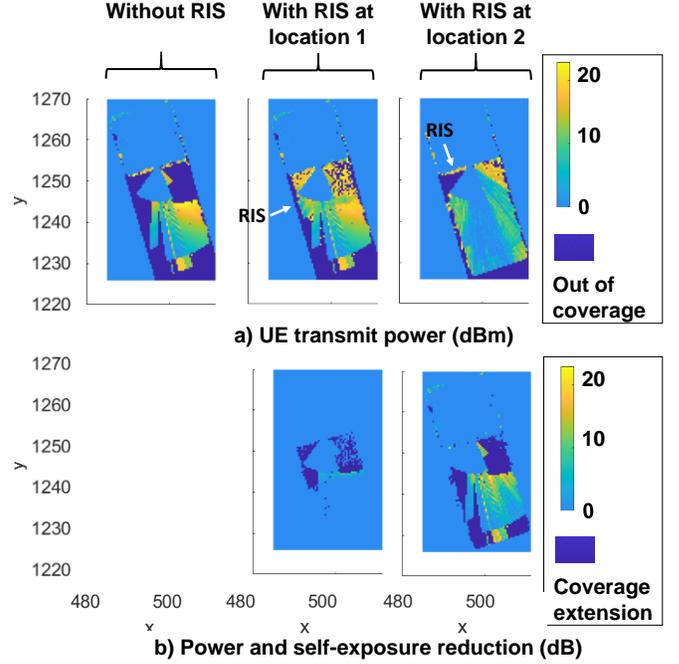

Fig. 5. Maps of a) UE transmit power, and b) UE transmit power reduction thanks to RIS.

## C. Reduced Self-Exposure Area

In this section, we assess numerically the UE transmit power (and thus the user self-exposure) improvement brought by a RIS for a target service and visualize the "reduced EMF exposure area".

In Fig. 5-a), maps of the UE transmit power is plotted for three different scenarios (without RIS, with RIS at location 1, and RIS at location 2). First of all, in room 2 and outside the building, the UE can use the minimum transmit power and still reach the target SNR, as the propagation conditions are excellent. In some locations inside room 1 and all locations behind the building, the UE is out of coverage, due to the multiple walls between the UE and the BS. Furthermore, we observe that the introduction of RIS has two effects: 1) it reduces the transmit power (and therefore the self-exposure) in locations that were already covered without RIS and 2) it reduces the size of the area out of coverage, i.e. in other terms, it extends the coverage area. Finally, we observe that the RIS location 2 is better than location 1, as expected based on the observations of Section B.

In Fig. 5-b), we plot the map of the UE transmit power (and therefore the self-exposure) reduction. We can observe that when deployed in location 1, the RIS EMF exposure reduction is much lower than when the RIS is in location 2 (where it can even reach the maximum of 23 dB of reduction).

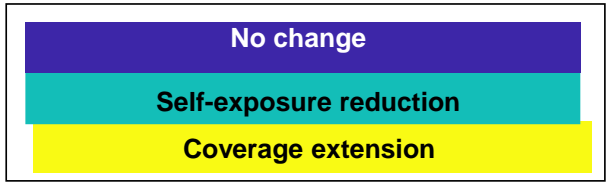

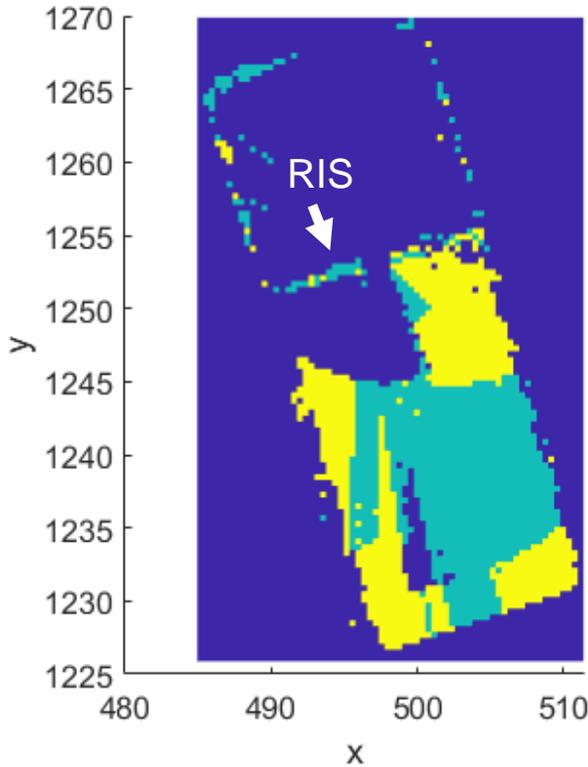

Fig. 6. Reduced self-exposure area, extended coverage area and area with no change.

Finally, in Fig. 6, we plot, for RIS in location 2, the areas where the following differences are observed when RIS has been introduced: 1) area where no change is observed; 2) area where the self-exposure is reduced (i.e. the "reduced EMF exposure area"); 3) area where the coverage is extended.

In a conclusion, an operator could potentially create areas of reduced EMF exposure, under the condition that the RIS may be deployed in the right location. Such location is very different from the optimum location for an active relay and strongly depends on the propagation between the RIS and the BS.

## IV. CONCLUSION

In this paper, we show that the self-exposure of users due to them using their smartphones to transmit data can be reduced in some areas, named "reduced EMF exposure area" thanks to the introduction of a reconfigurable intelligent surface. Simulations show that the placement method of a RIS cannot be similar to the deployment methods for an active relay. Indeed, in the chosen simulated environment, which is challenging regarding self-exposure, it is better to deploy the RIS in one room to reduce the self-exposure in another one. Next studies will further analyze RIS deployment methods. As current simulations have been held with a "giant" 10,000 element RIS to determine the upper bound performance, further studies will take into account RIS with smaller sizes and optimal deployment methods of such RISs.

ACKNOWLEDGMENT

This work was partially conducted within the framework of the European Union's Horizon 2020 research and innovation project RISE-6G under EC Grant 101017011.

REFERENCES

[1] L. Chiaraviglio, A. Elzanaty and M. -S. Alouini "Health Risks Associated With 5G Exposure: A View From the Communications Engineering Perspective," in *IEEE Open J. ComSoc*, vol. 2, pp. 2131-2179, 2021.

[2] M. D. Renzo, et al. "Smart Radio Environments Empowered by Reconfigurable AI Meta-Surfaces: an Idea whose Time has Come," *EURASIP J. on Wireless Comm. and Net.*, May 2019.

[3] E. Basar et al. "Wireless Communications through Reconfigurable Intelligent Surfaces," IEEE Access, vol. 7, pp. 116753-116773, 2019.

[4] C. Huang et al., "Reconfigurable Intelligent Surfaces for Energy Efficiency in Wireless Communication," in *IEEE Transactions on Wireless Communications,* vol. 18, no. 8, pp. 4157-4170, Aug. 2019.

[5] N. Awarkeh et al. "Electro-Magnetic Field (EMF) aware beamforming assisted by Reconfigurable Intelligent Surfaces," in *Proc. IEEE 22nd International Workshop on Signal Processing Advances in Wireless Communications (SPAWC)*, pp. 541-545, Sept 2021.

[6] F. Rusek et al., "Scaling up MIMO: Opportunities and Challenges with Very Large Arrays," IEEE Sig. Proc. Mag., vol. 30, pp. 40-60, Jan. 2013.

[7] H. Ibraiwish et al. "EMF-Aware Cellular Networks in RIS-Assisted Environments," in *IEEE Com. Letters*, vol. 26, pp. 123-127, Jan. 2022.

[8] A. Zappone, M. Di Renzo "Energy Efficiency Optimization of Reconfigurable Intelligent Surfaces with Electromagnetic Field Exposure Constraints," submitted to *Signal Proc. Letters*, Jan. 2022.

[9] E. C. Strinati et al., "Wireless Environment as a Service Enabled by Reconfigurable Intelligent Surfaces: The RISE-6G Perspective," in *Proc. 2021 Joint EuCNC and 6G Summit*, 2021, pp. 562-567.

[10] E. C. Strinati et al., "Reconfigurable, Intelligent, and Sustainable Wireless Environments for 6G Smart Connectivity," in IEEE Communications Magazine, vol. 59, no. 10, pp. 99-105, October 2021.

[11] George C. Alexandropoulos et al. "Smart Wireless Environments Enabled by RISs: Deployment Scenarios and Two Key Challenges," accepted to *2022 EUCNC & 6G Summit conference*, Grenoble, June 2022.

[12] A. Albanese et al. "RIS-Aware Indoor Network Planning: The Rennes Railway Station Case," accepted to *2022 IEEE ICC,* 2022.

[13] R. Fara et al. "Reconfigurable Intelligent Surface-Assisted Ambient Backscatter Communications – Experimental Assessment, " in *Proc. 2021 IEEE ICC Workshops*, 2021, pp. 1-7.

[14] R. Fara et al, "A Prototype of Reconfigurable Intelligent Surface with Continuous Control of the Reflection Phase," in *IEEE Wireless Communications*, vol. 29, no. 1, pp. 70-77, February 2022.

[15] 3rd Generation Partnership Project; Technical Specification Group Radio Access Network; Evolved Universal Terrestrial Radio Access; Radio Frequency requirements for LTE Pico Node B  (Release 16).

[16] 5G; NR; User Equipment radio transmission and reception; Part 2: Range 2 Standalone (3GPP TS 38.101-2 version 15.2.0 Release 15).